\documentstyle[11pt]{article}
\newtheorem{theorem}{Theorem}

\newcommand{\eqdf}{\stackrel{\triangle}{=}}

\newcommand {\reals} {{\rm I\!R}}

\newcommand {\bu} {\mbox{\boldmath $u$}}

\newcommand {\bE} {\mbox{\boldmath $E$}}

\newcommand {\bU} {\mbox{\boldmath $U$}}

\newcommand{\calC}{{\cal C}}

\newcommand{\calE}{{\cal E}}

\newcommand{\calH}{{\cal H}}
\newcommand{\calI}{{\cal I}}

\newcommand{\calU}{{\cal U}}

\newcommand{\calX}{{\cal X}}

\begin{document}
\thispagestyle{empty}
\setcounter{page}{1}
\title{On Optimum Parameter Modulation--Estimation From a Large Deviations
Perspective}
\author{Neri Merhav}

\date{}
\maketitle

\begin{center}
Department of Electrical Engineering \\
Technion - Israel Institute of Technology \\
Technion City, Haifa 32000, ISRAEL \\
E--mail: {\tt merhav@ee.technion.ac.il}\\
\end{center}
\vspace{1.5\baselineskip}
\setlength{\baselineskip}{1.5\baselineskip}

\begin{abstract}
We consider the problem of jointly optimum modulation and estimation of a
real--valued random parameter, conveyed over an additive white Gaussian
noise (AWGN) channel, where the performance metric is the large deviations
behavior of the estimator, namely, the exponential decay rate 
(as a function of the observation time) of the
probability that the estimation error would exceed a certain threshold.
Our basic result is in providing an exact characterization of the fastest
achievable exponential decay rate, among all possible modulator--estimator
(transmitter--receiver) pairs, where the modulator is limited only in the
signal power, but not in bandwidth. This exponential rate turns out to be
given by the reliability function of the AWGN channel.
We also discuss several ways to achieve this optimum performance, and one of
them is based on quantization of the parameter, followed by optimum channel
coding and modulation, which gives rise to a separation--based transmitter, if
one views this setting from the perspective of joint source--channel coding.
This is in spite of the fact that, in general, when error exponents are
considered, the source--channel separation theorem does not hold true.
We also discuss several observations,
modifications and extensions of this result in several directions,
including other
channels, and the case of 
multidimensional parameter vectors. One of our findings
concerning the latter, is that there is an abrupt threshold effect in the
dimensionality of the parameter vector: below a certain critical dimension,
the probability of excess estimation error may still decay exponentially, but
beyond this value, it must converge to unity.\\

\noindent
{\bf Index Terms:} Parameter estimation, modulation, AWGN, threshold effect, large
deviations, reliability function, error exponents.
\end{abstract}

\clearpage

\section{Introduction}

The rich literature on parameter estimation includes a large 
variety of Bayesian and non--Bayesian lower bounds on
the mean square error (MSE) in estimating parameters from signals
corrupted by an additive white Gaussian noise (AWGN) channel, as well as other
channels (see, e.g., the introductions of \cite{BSET97},
\cite{BE09}, \cite{Weiss85} 
for overviews 
on these bounds). Most of these bounds are
amenable to calculation for
a given form of dependence of the transmitted signal upon the parameter,
i.e., a given modulator, and 
therefore they may give insights concerning optimum
estimation for this specific modulator.  
They may not, however, lend themselves easily to the derivation of
{\it universal} lower bounds, namely, lower bounds that
depend neither on the modulator nor on the estimator, which are relevant
when both optimum modulators and optimum
estimators are sought. Two exceptions to this rule 
(although usually, not presented as such)
are families of bounds
that stem from generalized data processing theorems (DPT's) \cite{p146}, \cite{ZZ75}, \cite{ZZ73},
and bounds based on hypothesis testing 
considerations \cite{CZZ75}, \cite{ZZ69b}.

Consider, for a example, a random parameter $U$, uniformly distributed
across the unit interval, which is to be conveyed across the AWGN channel with
spectral density $N_0/2$, transmission power $S$, 
and no bandwidth limitation. 
Using the classical DPT, 
one views the random parameter $U$
as a ``source'' and the MSE of an arbitrary estimator, $\bE(\hat{U}-U)^2$, as
the average distortion $D$, and then derives a lower bound on $D$ from the
inequality $R(D)\le CT$, where $R(D)$ is the rate--distortion function of $U$,
$T$ is the
transmission time, and $C$ is the channel capacity, 
which for the AWGN
with unlimited bandwidth, is given by
$C=S/N_0$. 
Now, $R(D)$ is
not known to have a closed--form expression in this case, but it can be
further lower bounded by the Shannon lower bound (see, e.g.,
\cite[Sect.\ 4.6, p.\ 101]{Gray90}):
\begin{equation}
R(D)\ge h(U)-\frac{1}{2}\ln (2\pi e D)=-\frac{1}{2}\ln (2\pi e D), 
\end{equation}
where $h(U)=0$ is the differential entropy of $U$.
This readily leads to the universal lower bound 
$\bE(\hat{U}-U)^2\ge \frac{1}{2\pi e}e^{-2CT}=\frac{1}{2\pi e}e^{-2\calE/N_0}$,
where $\calE=ST$ is the signal energy.
It turns out that this lower bound is not tight. In \cite{ZZ75}, it was shown
that DPT's pertaining to generalized information
measures, yield a tighter
universal lower
bound that decays (as $T\to\infty$) like $e^{-CT}$. In \cite{p146}, this bound was further
improved, by another generalized DPT, 
to behave like $e^{-2CT/3}$, and then 
yet further improved
to $e^{-CT/2}$, using a universal lower bound based 
on signal detection considerations,
in the spirit of the Ziv--Zakai bound \cite{ZZ69b} and the Chazan--Zakai--Ziv
bound \cite{CZZ75}. 

Concerning upper bounds,
it turns out that it is possible to achieve an 
MSE with an exponential decay rate of the
order of $e^{-CT/3}$, which is quite close to the latter lower bound, but
there is still some gap.
As is shown in \cite[Chap.\ 8]{WJ65}, by using frequency
position modulation (FPM)
with central frequency and bandwidth that both grow
like $e^{RT}$, where $R>0$ is a fixed design parameter, the MSE of the maximum likelihood
(ML) estimator turns out to be composed of two terms: 
a ``small--error'' term (or the ``weak noise performance'' in the terminology of
\cite{WJ65}),
that behaves essentially like the Cram\'er--Rao bound, and which is proportional to
$e^{-2RT}$, and an anomalous error term (gross error due to the threshold
effect) of the exponential order
of $e^{-E(R)T}$, where $E(R)$ is the reliability 
function of the AWGN, given by
\begin{equation}
\label{ER}
E(R)=\left\{\begin{array}{ll}
\frac{C}{2}-R & 0\le R\le \frac{C}{4}\\
(\sqrt{C}-\sqrt{R})^2 & \frac{C}{4}\le R \le C\end{array}\right.
\end{equation}
The optimum trade-off between these two terms 
is achieved for $R=C/6$, where they have the
same exponential rate, $e^{-CT/3}$ (see also \cite{p137}).
Similar things can be
said about pulse position modulation (PPM) with exponentially growing
bandwidth \cite{p137}. Yet another modulation scheme is based on simply
quantizing the parameter $U$ into one of $M=e^{RT}/2$ evenly spaced points 
in its interval (which are
then far apart by $2e^{-RT}$) and then
assigning, to each one of these points, one out of $M$ orthogonal signals with
energy $\calE$ (see \cite{NVW12} for an analogue for the Poisson channel).
Here, the MSE has the same two exponential terms as before, but
now the first term, $e^{-2RT}$, is the contribution of the quantizer
to the MSE and the second term, $e^{-E(R)T}$, is the contribution of channel
decoding errors.

The quest for closing (or at least, further reducing) 
the gap between the best known lower bound,
$e^{-CT/2}$, and the upper bound, 
$e^{-CT/3}$, remains unsatisfied at present.
This challenge has, unfortunately, defied our best efforts thus far.
We conjecture that it is the lower bound that is to be ``blamed'' for this
gap, i.e., we believe
that the above--mentioned modulation schemes are
essentially optimal but there is room for further improvement 
of the lower bound 
that has not been exploited yet.

In this paper, instead of focusing on the MSE as our performance metric, we
adopt a large deviations performance metric: We seek optimum modulation and
estimation schemes in the sense of maximizing the exponential rate
of decay of the probability that the estimation error $|\hat{U}-U|$ would
exceed a given threshold. Motivated by the above discussion, we can afford
to set this threshold to be exponentially decaying with $T$, i.e., $e^{-RT}$,
where $R> 0$ is a parameter whose value can be chosen freely in some range.
More precisely, our asymptotic figure of merit for modulation--estimation is
\begin{equation}
\label{erdef}
E^*(R)=\limsup_{T\to\infty}\left[-\frac{1}{T}\log\inf\mbox{Pr}\left\{|\hat{U}-U|>
e^{-RT}\right\}\right],
\end{equation}
where the infimum is over all modulator--estimator pairs with power
$S$, and where we remind the reader that
$\limsup_{T\to\infty}f(T)$, for a continuous--valued variable $T$ (as opposed
to a sequence $\{T_n\}$), is defined as $\lim_{T\to\infty}\sup_{T'\ge T}f(T')$.

Our basic result (asserted and proved in Section 2) is that
the $\limsup$ in eq.\ (\ref{erdef}) is equal to the corresponding
$\liminf$ (and hence can be replaced by $\lim$) and their common value
has an exact characterization given by
$E^*(R)=E(R)$, 
where $E(R)$ is as
in (\ref{ER}).
All three modulation schemes mentioned above, together with ML
estimation, achieve this performance and hence are asymptotically optimum
in the above sense.\footnote{The fact that 
exponentially small error thresholds are
exceeded with exponentially small probabilities is rather remarkable. It is
thanks to the fact that the modulator is subjected to optimization. By
contrast, for
amplitude modulation (AM), where the estimation 
error of the ML estimator has variance
$N_0/(2\calE)=1/(2CT)$,
we have $\mbox{Pr}\{|\hat{U}-U|>e^{-RT}\}=2Q(e^{-RT}\sqrt{2CT})\to 1$ for
every $R>0$, and only for $R=0$ this probability decays exponentially.}

Beyond the fact that the large deviations performance metric has already been
addressed in estimation theory (see, e.g.,
\cite{KK86}, \cite[p.\ 4]{Lehmann83},
\cite{Sherman58}, \cite[p.\ 54]{vanTrees68}, \cite[eq.\ (32)]{WZ69},
\cite[Sect.\ IV]{ZZ69b}),
a little thought suggests that
it is actually natural in this particular setting of wide--band waveform
communication, which exhibits threshold effects and anomalies. 
The reason is that it makes a clear distinction between
`small' errors, of the order of $e^{-RT}$ (``allowed'' under this
metric), and gross errors, whose probabilistic weight is
$e^{-E(R)T}$ at best.\footnote{Typically, in the case of anomaly, the estimate
$\hat{U}$ falls in a random point away from $U$, and so, it makes sense
to assign to all gross error events the same cost, as is done by the proposed metric.}
A distinction in the same spirit (but not quite the same) was offered also in 
\cite[Sect.\ 8.4]{WJ65}, where it was shown that a non--anomalous MSE of about $e^{-2CT}$ is
the best that can be achieved (and again, by the same schemes)
under the constraint that the probability of anomaly
tends to zero. This has the flavor of our result for $R\approx C$, but here,
we expand the spectrum of trade-offs to 
the entire range $0\le R\le C$.
For $R> C$, the error exponent
vanishes in the strong sense, 
i.e., not only does the probability of the undesired
error event cease to decay exponentially, it actually tends to unity. In that
sense, the threshold effect is manifested in a clear way. 

Having hopefully convinced the reader that the large deviations performance
criterion is reasonable in the waveform communication setting
considered here, there
is considerable room for the 
speculation that this may not be the case with the
MSE criterion,
despite its popularity. 
The difficulty in capturing the
threshold effect and in closing gaps between upper and lower bounds in this
setting, as
discussed above, may be attributed to the fact that the MSE does not
distinguish between the small errors and the anomalous errors, which are
so different in nature. Comments in the same spirit 
are made also in \cite[p.\ 633, central paragraph]{WJ65}.

We discuss several observations and implications
of the above described basic result 
(Section 3) and several extensions (Section 4),
including other channels, variable power,
and the case of a multidimensional parameter vector
$\bU=(U_1,\ldots,U_d)$. In the vector case, our error exponent criterion becomes
\begin{equation}
E^*(R_1,\ldots,R_d)=\limsup_{T\to\infty}\left[-\frac{1}{T}\log\inf\mbox{Pr}
\left(\bigcup_{i=1}^d\left\{|\hat{U}_i-U_i|>
e^{-R_iT}\right\}\right)\right],
\end{equation}
where our earlier characterization, in terms of the reliability function,
extends to
\begin{equation}
E^*(R_1,\ldots,R_d)=E(R_1+R_2+\ldots+R_d).
\end{equation}
One of the conclusions of this result
is that there is an abrupt threshold effect in the
dimensionality of the parameter vector: below a certain critical dimension,
the probability of excess estimation error may still decay exponentially, but
beyond this value, it must converge to unity. We also discuss several other
implications of our results.

As a closing remark, we should point out that the criterion of
excess estimation error probability was briefly discussed also in \cite[Section
IV]{ZZ69b}, where a lower bound was given in terms of the error probability of
an $M$--ary detection problem with optimum signaling. 
This is similar to the line of thought here, however, there are
several differences:
(i) We consider a Bayesian setting where $U$ is a random
variable, as opposed to the worst--case excess error probability,
$\max_u \mbox{Pr}\{\mbox{excess estimation error}|u\}$. (ii) We allow
an arbitrary modulator, rather than focusing on PPM specifically. (iii) We
allow an exponentially vanishing error threshold, $e^{-RT}$ (as opposed to
a fixed threshold in \cite{ZZ69b}, corresponding to $R=0$) and explore the entire spectrum of
trade-offs between $R$ and the excess estimation error exponent, which
in turn is intimately related to the reliability function, $E(R)$.
(iv) As described in the previous paragraph, we also expand the scope in
several directions, like the multidimensional case and other channels. We also
provide some insights from the perspectives 
of the threshold effect as well as joint source--channel coding
and the separation theorem.

\section{Problem Formulation and the Basic Result}

Consider the signal model
\begin{equation}
y(t)=x(t,u)+n(t),~~~~t\in [0,T)
\end{equation}
where $x(t,u)$ is a waveform with power $S$,
which is parametrized by $u\in \calU\subseteq \reals$, 
and where $n(t)$ is AWGN with two--sided power spectral
density $N_0/2$. 
Considering an arbitrary representation of $x(t,u)$ as a
linear combination of 
orthonormal basis functions, then due to the power
limitation, the length of the curve (locus) drawn by the vector of
coefficients of this representation, $\{a_k(u),~k=1,2,\ldots\}$, 
as $u$ exhausts $\calU$, must be finite
(and in fact, no larger than $e^{CT}$ \cite[Chap.\ 8]{WJ65}) in order to
keep the anomalous error vanishingly small. It therefore makes sense to assume that
$\calU$ is a finite interval, which without loss of essential generality, will
be taken to be the interval $[-1/2,+1/2)$, as any other interval can be
obtained under re-parametrization using a simple affine transformation.

An estimator of $u$ is any measurable mapping from $\{y(t),~0\le t <T\}$ into
$\calU$.
In order to avoid limitations on the class of estimators (e.g., unbiased
estimators, etc.), we
adopt the Bayesian setting, i.e., we assume that $u$ is a realization of a random
variable $U$, uniformly distributed over $[-1/2,+1/2)$. The uniform prior is
assumed merely for convenience and it expresses the fact that no value of $u$
has any preference {\it a-priori}. Any other prior, which is bounded away from
zero and infinity, can be used as well.

A modulator with power $S$ is a mapping from $\calU$ into a family of waveforms
$\{x(t,\cdot),~0\le t<T\}$, whose power is exactly\footnote{In Subsection
\ref{varpower}, we relax the restriction that the power would be exactly $S$
for all $u$, and we allow instead the power $S(u)$ to vary with $u$, but we keep
an average power constraint, $\bE\{S(U)\}\le S$.}
$S$, i.e.,
\begin{equation}
\frac{1}{T}\int_0^T\mbox{d}t\cdot x^2(t,u)=S
\end{equation}
for all $u\in\calU$. No bandwidth limitations are imposed on the
waveforms in this family.

For a given $R>0$,
we are interested in characterizing the best achievable excess estimation
error exponent 
\begin{equation}
E^*(R)=\limsup_{T\to\infty}\left[-\frac{1}{T}\log\inf\mbox{Pr}\left\{|\hat{U}-U|>
e^{-RT}\right\}\right],
\end{equation}
where the infimum is over all
modulator--estimator pairs as defined as above.

We first provide a lower bound on the excess estimation error probability,
that leads directly to a converse theorem concerning $E^*(R)$.
\begin{theorem}
Consider the AWGN channel with noise power spectral density $N_0/2$.
Let $R > 0$ be given and let $\epsilon > 0$ be arbitrarily small.
For every modulator with power $S$ and every estimator $\hat{U}$:
\begin{equation}
\mbox{Pr}\left\{|\hat{U}-U|> e^{-RT}\right\}\ge (1-e^{-\epsilon
T})\exp\{-T[E(R-\epsilon)+o(T)]\},
\end{equation}
where $E(R)$ is the reliability function of the AWGN, defined as
in eq.\ (\ref{ER}) and
where $o(T)$ designates a quantity that tends to zero as $T\to\infty$.
Consequently, 
\begin{equation}
E^*(R)\le E(R).
\end{equation}
\end{theorem}

While the lower bound in Theorem 1 
applies, in principle, for every $\epsilon > 0$,
quite obviously, for $T\to\infty$, the tightest lower bound is obtained as
$\epsilon\to 0$, which yields an exponential decay rate of $E(R)$.\\

\noindent
{\it Proof.}
The proof is in the spirit of the derivation of the Ziv--Zakai bound
\cite{ZZ69b} and the Chazan--Zakai--Ziv bound \cite{CZZ75}, but with
$M$ hypotheses (rather than $2$), where $M$ is exponentially large.
Consider a given estimator $\hat{U}$ of $U$ and a given modulator
$\{x(t,\cdot),~0\le t< T\}$ with power $S$. For
a given $u\in [-1/2,+1/2)$ and $\Delta > 0$, let $P_e(u,\Delta)$ denote the probability of error
of the optimum (ML) detector for deciding among the $M$ equiprobable hypotheses
$$\calH_i:~~y(t)=x(t,u+i\Delta)+n(t),~~~i=0,1,\ldots,M-1$$
where it is assumed that $u$ and $\Delta$ are such
that $u+i\Delta$, $i=0,1,\ldots,M-1$, are all 
in $[-1/2,+1/2)$. First, it is argued that
\begin{eqnarray}
P_e(u,\Delta)&\le&\frac{1}{M}\left[\mbox{Pr}\left\{\hat{U}-U>
\frac{\Delta}{2}\bigg| U=u\right\}+\right.\nonumber\\
& &\sum_{i=1}^{M-2}\mbox{Pr}\left\{|\hat{U}-U|>
\frac{\Delta}{2}\bigg| U=u+i\Delta\right\}+\nonumber\\
& &\left.\mbox{Pr}\left\{\hat{U}-U<
-\frac{\Delta}{2}\bigg| U=u+(M-1)\Delta\right\}\right].
\end{eqnarray}
To see why this is true, observe that the r.h.s.\ can be interpreted as
the probability of error of a suboptimum $M$--ary detector that is based on
first estimating $U$ by $\hat{U}$ and then deciding on the hypothesis
$\calH_i$ whose corresponding grid point $u+i\Delta$ 
is nearest to $\hat{U}$. Next, we further upper bound the first
and the last terms of the r.h.s.\ by $\mbox{Pr}\{|\hat{U}-U|>\Delta/2|U=u\}$
and $\mbox{Pr}\{|\hat{U}-U|>\Delta/2|U=u+(M-1)\Delta\}$, respectively,
which yields
\begin{equation}
P_e(u,\Delta)\le \frac{1}{M}\sum_{i=0}^{M-1}\mbox{Pr}\left\{|\hat{U}-U|>
\frac{\Delta}{2}\bigg| U=u+i\Delta\right\}.
\end{equation}
Integrating both sides over
$u$, we get
\begin{eqnarray}
\label{zz}
& &\int_{-1/2}^{+1/2-(M-1)\Delta}\mbox{d}u\cdot P_e(u,\Delta)\nonumber\\&\le&
\int_{-1/2}^{+1/2-(M-1)\Delta}\mbox{d}u\cdot\frac{1}{M}
\sum_{i=0}^{M-1}\mbox{Pr}\left\{|\hat{U}-U|>
\frac{\Delta}{2}\bigg| U=u+i\Delta\right\}\nonumber\\
&=&\frac{1}{M}\sum_{i=0}^{M-1}
\int_{-1/2}^{+1/2-(M-1)\Delta}\mbox{d}u\cdot\mbox{Pr}
\left\{|\hat{U}-U|>
\frac{\Delta}{2}\bigg| U=u+i\Delta\right\}\nonumber\\
&=&\frac{1}{M}\sum_{i=0}^{M-1}\int_{-1/2+i\Delta}^{+1/2-(M-1)\Delta+i\Delta}
\mbox{d}u\cdot\mbox{Pr}\left\{|\hat{U}-U|>
\frac{\Delta}{2}\bigg| U=u\right\}\nonumber\\
&\le&\frac{1}{M}\sum_{i=0}^{M-1}\int_{-1/2}^{+1/2}
\mbox{d}u\cdot\mbox{Pr}\left\{|\hat{U}-U|>
\frac{\Delta}{2}\bigg| U=u\right\}\nonumber\\
&=&\int_{-1/2}^{+1/2}\mbox{d}u\cdot\mbox{Pr}\left\{|\hat{U}-U|\ge
\frac{\Delta}{2}\bigg| U=u\right\}\nonumber\\
&=&\mbox{Pr}\left\{|\hat{U}-U|>
\frac{\Delta}{2}\right\}.
\end{eqnarray}
Now, let $\Delta=2e^{-RT}$ and $M=e^{(R-\epsilon)T}/2+1$.
Then, it is well known (see, e.g., \cite[p.\
168, eq.\ (3.6.26) and Section 3.8]{VO79},
\cite[p.\ 383, eqs.\ (8.2.49),
(8.2.50)]{Gallager68}, \cite[pp.\ 345, eq.\ (5.106c)]{WJ65}) that
\begin{equation}
P_e(u,\Delta)\ge 
e^{-T[E(R-\epsilon)+o(T)]},
\end{equation}
which, when substituted into the left--most side of (\ref{zz}), readily gives
\begin{eqnarray}
\mbox{Pr}\left\{|\hat{U}-U|> e^{-RT}\right\}&\ge&
\int_{-1/2}^{+1/2-(M-1)\Delta}\mbox{d}u\cdot
e^{-T[E(R-\epsilon)+o(T)]}\nonumber\\
&=&[1-(M-1)\Delta]e^{-T[E(R-\epsilon)+o(T)]}\nonumber\\
&=&(1-e^{-\epsilon T})e^{-T[E(R-\epsilon)+o(T)]},
\end{eqnarray}
completing the proof of Theorem 1. $\Box$

Our next theorem, Theorem 2, provides a compatible achievability result.
\begin{theorem}
Consider the AWGN channel with noise power spectral density $N_0/2$ and
let $R > 0$ be given. Then
there exists a modulator with power $S$ and an estimator $\hat{U}$
for which
\begin{equation}
{Pr}\left\{|\hat{U}-U|> e^{-RT}\right\}\le
e^{-E(R)T}.
\end{equation}
Consequently, the $\limsup$ in eq.\ (\ref{erdef}) is
equal to the $\liminf$ (i.e., the limit exists) and 
\begin{equation}
E^*(R)=E(R).
\end{equation}
\end{theorem}

\noindent
{\it Proof.}
We first describe the modulator and estimator. 
Assume, without essential loss of generality, that $e^{RT}/2$ is integer (otherwise,
alter the value of $R$ slightly to make it such).
The modulator first quantizes
the parameter $u$ to the nearest point in the grid
$\{-1/2+e^{-RT},-1/2+3e^{-RT},-1/2+5e^{-RT},\ldots,1/2-e^{-RT}\}$. This grid,
which consists of
$M=e^{RT}/2$ points, is mapped into a set of $M$ orthogonal signals, each with power
$S$. Let $i(u)$ denote the index of the grid point nearest to $u$ and
let $x_i(t)$ be the signal corresponding to the $i$--th grid point,
$i=1,2,\ldots,M$. Then the modulator is defined by 
\begin{equation}
x(t,u)=x_{i(u)}(t).
\end{equation}
Let $\hat{i}$ denote the output of the ML decoder for the signal set
$\{x_i(t)\}_{i=1}^M$, namely,
\begin{equation}
\hat{i}=\mbox{argmax}_{1\le i\le M}\int_0^Tx_i(t)y(t)\mbox{d}t.
\end{equation}
Then, the estimator $\hat{u}$ is defined as the corresponding grid point,
i.e., 
\begin{equation}
\hat{u}=-\frac{1}{2}+(2\hat{i}-1)e^{-RT}.
\end{equation}
Clearly, for this 
particular modulator--estimator pair, the event $\{|\hat{U}-U|>
e^{-RT}\}$ implies $\hat{i}\ne i(U)$, namely, an error in
decoding the index $i$ of the transmitted signal $x_i(t)$. The probability of
excess estimation error is therefore upper bounded by
the probability of error for $M=e^{RT}/2$ orthogonal
signals, each with energy $\calE=ST$, which is well known
(see, e.g., \cite[p.\ 67, eq.\ (2.5.16)]{VO79} or \cite[p.\
381, eqs.\ (8.2.43), (8.2.44)]{Gallager68}, \cite[pp.\ 344--345, eqs.\
(5.104)--(5.106b)]{WJ65}) to be upper bounded in turn by
$e^{-E[R-(\ln 2)/T]T}\le e^{-E(R)T}$.
This completes the proof of Theorem 2. $\Box$\\

\noindent
{\bf The Case $R=0$}

Theorems 1 and 2 refer to the case $R> 0$.
The case $R=0$ should be treated with caution as there is an inherent
discontinuity of the {\it operational} reliability function at $R=0$. 
As is well known, the operational reliability function for the infinite--bandwidth AWGN channel, 
which is defined as the asymptotic error
exponent of the optimum rate--$R$ code for this channel, agrees with $E(R)$,
given in (\ref{ER}), only for $R> 0$. Concerning the point $R=0$,
there is a difference between the strong sense of this assignment,
where the number of codewords $M$ is fixed (independent of $T$),
and the weak sense, where $M$ grows (but in a subexponential rate). This is because
for fixed $M$, the error exponent of the best signal set (the simplex signal set) is
determined by the minimum distance,
which depends on $M$ according to $d_{\min}=2M\calE/(M-1)$, 
where again, $\calE=ST$ is the
energy of all $M$ signals. The error probability of the optimum code then
decays according to $\exp[-T\frac{C}{2}\cdot\frac{M}{M-1}]$, which agrees with
$E(0)=C/2$ only when $M$ grows without bound. 

Correspondingly, there is a parallel
difference between the case
where the error threshold,
$\Delta/2$ (in the proof of Theorem 1) is {\it fixed}, as opposed to the weaker sense
where $\Delta$ is allowed to vanish as $T$ grows, but in a subexponential rate.
Theorems 1 and 2 hold for $R>0$, and the limit $R\to 0$ corresponds to the
weaker meaning. What can be said about the stronger meaning?
Repeating the proof of Theorem 1, but with a
zero--rate lower bound on $P_e(u,\Delta)$ \cite[p.\ 174, eqs.\
(3.7.2)--(3.7.5)]{VO79}, \cite[pp.\ 345, eq.\ (5.106c)]{WJ65},
we have (by choosing $M=\lfloor 1/\Delta\rfloor$)
\begin{equation}
\label{lr=0}
\mbox{Pr}\left\{|\hat{U}-U|> \frac{\Delta}{2}\right\}\ge
\frac{1}{2}(1+\Delta-\Delta\lfloor 1/\Delta\rfloor)\cdot
Q\left(\sqrt{\frac{\calE}{N_0}\cdot\frac{\lfloor 1/\Delta\rfloor}
{\lfloor 1/\Delta\rfloor-2}}\right).
\end{equation}
In the limit of $T\to\infty$, this lower bound is of the exponential order of
$$\exp\left\{-\frac{CT}{2}\cdot \frac{\lfloor 1/\Delta\rfloor}{\lfloor
1/\Delta\rfloor-2}\right\}.$$
As an upper bound we have, by a compatible upper bound on the probability of
error (see proof of Theorem 2), the following:
\begin{equation}
\label{ur=0}
\mbox{Pr}\left\{|\hat{U}-U|> \frac{\Delta}{2}\right\}\le
\left(\lceil 1/\Delta\rceil-1\right)\cdot
Q\left(\sqrt{\frac{\calE}{N_0}\cdot\frac{\lceil 1/\Delta\rceil}
{\lceil 1/\Delta\rceil-1}}\right),
\end{equation}
which simply follows from the union bound on the probability of error in the
detection of one out of $M=\lceil 1/\Delta\rceil$ simplex signals with energy
$\calE=ST$. Here, the exponential behavior is according to
$$\exp\left\{-\frac{CT}{2}\cdot \frac{\lceil 1/\Delta\rceil}{\lceil
1/\Delta\rceil-1}\right\}.$$
While there is a gap in the error exponents for every finite $\Delta$, this
gap vanishes as $\Delta\to 0$, thus the best achievable asymptotic value of
$$\lim_{\Delta\to 0}\lim_{T\to\infty}\left[-\frac{\ln \mbox{Pr}\{|\hat{U}-U|>
\Delta/2\}}{T}\right]$$ is still $E(0)=C/2$.

In this context of large deviations for fixed $\Delta$, it is appropriate to
mention also the relation
with the MSE criterion: The two criteria are
easily related via the identity
\begin{equation}
\label{mseld}
\bE(\hat{U}-U)^2=
2\int_0^1\mbox{d}\Delta\cdot\Delta\cdot\mbox{Pr}\{|\hat{U}-U|\ge\Delta\},
\end{equation}
and so, the MSE can be lower bounded via any lower bound on
$\mbox{Pr}\{|\hat{U}-U|\ge\Delta\}$ for all $\Delta$ in the appropriate range,
which is exactly the line of thought that guides the Chazan--Zakai--Ziv bound
\cite{CZZ75} for two hypotheses. Here, as we consider $M$ hypotheses rather
than two,\footnote{Here, two hypotheses correspond to antipodal signals, rather than
orthogonal signals, and hence lead to non--tight exponential error bounds
with a loss of 3dB.}
and $M$ is exponentially large, the integration range of $\Delta$ in
the corresponding lower bound, where the integrand is $P_e(u,\Delta)$, must be
limited to the interval $(0,1/(M-1)]$, as otherwise, some grid points
$\{u+i\Delta\}$ (in the proof of Theorem 1), 
would fall outside the interval $[-1/2,+1/2)$. This limitation on the
range of $\Delta$ causes the resulting lower bound on the MSE to be relatively
weak. One of the main points in this paper is that 
by considering the large deviations performance as our figure of merit
in the first place,
we actually avoid the need to integrate over $\Delta$ altogether. 
An interesting open question, in this context, is whether it is possible to
devise a modulator--estimator pair, which would be independent of $\Delta$,
but yet achieve asymptotically optimum large deviations performance for all $\Delta$ in
the interesting range. Such an estimator may also achieve
asymptotically optimum MSE, in view of eq.\ (\ref{mseld}).

\section{Discussion}

In this section, we pause to discuss a few observations, implications, and modifications 
of Theorems 1 and 2.

\subsection{Strong Converse and the Threshold Effect}

The case $R=0$, discussed in Section 2, is one interesting extreme of the
range of $R$. The other extreme is the point $R=C$, where $E(R)$
vanishes. Here, due to the strong converse to the channel coding theorem,
$E(R)$ vanishes in the strong sense for $R> C$, namely, the probability
of error tends to unity. Owing to the proof of Theorem 1, 
the large deviations
estimation performance criterion, considered in this paper, `inherits' this strong
converse, and then the probability of excess estimation error tends to unity
as $T\to \infty$, for $R > C$. This means an abrupt threshold effect in the
limiting probability of
excess error, from $0$ to $1$, as $R$ crosses $C$.

\subsection{Achievability by Other Schemes}

As mentioned in the Introduction,
alternative achievability proofs are possible
by analyzing FPM and PPM systems. The FPM modulator (see, e.g., \cite{WJ65})
is defined as follows:
\begin{equation}
x(t,u)=\sqrt{2S}\cos[2\pi(f_0+u\cdot\Delta f)t], 
\end{equation}
where in our case, both the central frequency
$f_0$ and the frequency offset
$\Delta f$ ($\Delta f \ll f_0$) are taken to be proportional to
$e^{RT}$. For the ML estimator
$\hat{U}$,
in this case, $\mbox{Pr}\{|\hat{U}-U|>  e^{-RT}\}$
is the probability of anomaly, which is essentially $e^{-E(R)T}$
(see \cite[eqs.\ (8.175a)--(8.175c)]{WJ65}).

Another good modulator for our purpose is PPM,
where
\begin{equation}
x(t,u)=s[t-(u+1/2)(T-\tau)], 
\end{equation}
$s[\cdot]$ being a pulse
whose support is $[0,\tau]$, where $\tau$
is proportional to $e^{-RT}$ (and hence
the bandwidth is proportional to $e^{RT}$), and again,
the large deviations event in question is the anomaly event (see, e.g.,
\cite{p137}, \cite{ZZ69a} for more details).

\subsection{Relation to Bounds on Moments of the Estimation
Error}

The combination of Theorem 1 with Chebyshev's inequality,
\begin{equation}
\mbox{Pr}\left\{|\hat{U}-U|> e^{-RT}\right\}
\le\frac{\bE(\hat{U}-U)^2}{e^{-2RT}}
\end{equation}
yields the following lower bound on the MSE
\begin{equation}
\bE(\hat{U}-U)^2\ge (1-e^{-\epsilon T})e^{-T[E(R-\epsilon)+2R+o(T)]},
\end{equation}
which is tightest for $R=\epsilon\to 0$, as $T\to\infty$. Thus, the
MSE is lower bounded by an expression whose exponential order is
$e^{-E(0)T}=e^{-CT/2}$, as discussed in the Introduction (see also
\cite{p146}). The same comment applies, of course, to
more general moments of the
estimation error, $\bE|\hat{U}-U|^\alpha$, in the range $\alpha\ge 1$
(see also \cite[p.\ 388, Remark 1]{ZZ69b}).
For $0< \alpha < 1$, the best choice of $R$ is near $R=C/(1+\alpha)^2$
and the resulting lower bound is of the exponential
order of $\exp[-\alpha CT/(1+\alpha)]$.

\subsection{Relation to the Joint Source--Channel Excess Distortion Exponent}

Note that if we think of the random parameter 
$U$ as a source variable, and then the modulation--estimation problem
is considered as a joint source--channel coding problem, then
our conclusion from Theorem 2 is that {\it separate 
source- and channel coding is asymptotically optimum} in our setting:
In the modulation scheme analyzed in the proof of Theorem 2,
the transmitter
first uses a source encoder that quantizes the
parameter $U$, 
applying a simple 
uniform scalar
quantizer -- see also \cite{NVW12}, 
and then maps the quantized version of $U$ into a channel input
waveform using a good channel code. The same comment applies to the
case where the parameter is a vector $\bU=(U_1,\ldots,U_d)$, as will be
discussed in Subsection \ref{md}, where the source encoder will quantize 
each component $U_i$ individually.

It is interesting to contrast this with the results of Csisz\'ar 
\cite{Csiszar82} (see also \cite{Csiszar80}), 
where exponential rates of probabilities of excess
end--to--end distortion between a source vector and its reconstruction vector
were studied under a joint source--channel coding setting.\footnote{In other
words, instead of analyzing the performance of the communication system under the criterion of average
distortion, it was analyzed in \cite{Csiszar82} under the probability that the
block distortion would exceed a certain threshold in the large deviations
regime.}
In that work, it was 
argued that, in general, separate source- and channel coding 
is sub-optimum in the error exponent sense
(see discussion at the 
second to the last paragraph in the Introduction of \cite{Csiszar82} 
as well as in \cite[Introduction]{Csiszar80} and
\cite[Problem 5.16, pp.\ 534--535]{Gallager68}). 
The natural question that arises is how do these two (seemingly contradicting)
facts settle, if there is any contradiction. 
First, observe that 
there are
some differences between our setting and the one in
\cite{Csiszar82}:
\begin{enumerate}
\item 
In our setting, the source variable $U$ is a scalar, namely, 
it remains of ``block--length'' $1$,
when $T$ goes to infinity, whereas
in \cite{Csiszar82} the analogous quantities grow together 
with a fixed ratio (which is known as the
bandwidth expansion factor). 
Even in Subsection \ref{md}, where as mentioned earlier, 
we extend our setting to the case of a vector
parameter, 
$\bU=(U_1,\ldots,U_d)$, the dimension $d$ will be assumed
fixed while $T\to\infty$.
\item As another difference in the asymptotic regime, in our case,
the allowed distortion threshold decays exponentially, whereas in
\cite{Csiszar82} it is fixed.
\item For the AWGN with infinite bandwidth, the reliability function is
fully known, as opposed to that of a general DMC.
\end{enumerate}
Nonetheless, in spite of these differences,
our results can be understood in the framework of \cite{Csiszar82}.
It turns out that while in general, 
there is no separation theorem for error exponents,
the parameter modulation--estimation problem considered here is analogous
to a special case, where a separation theorem holds true 
for error exponents nevertheless.

To be more specific, Csisz\'ar's main result in \cite{Csiszar82} 
can be presented essentially as follows:
The best excess distortion exponent of joint source--channel coding is
upper bound by
\begin{equation}
e(D)=\min_R[F(D,R)+ E(R)], 
\end{equation}
where 
\begin{equation}
F(D,R)=\min_{\{Q':~R(D,Q')\ge R\}}
D(Q'\|Q) 
\end{equation}
is Marton's source coding (excess distortion) exponent
of the source $Q$ \cite{Marton74}, 
$R(D,Q')$ being the rate--distortion function of a source $Q'$,
and $E(R)$ is the reliability function of the channel.
Now, consider the source $Q^*$ that maximizes
$R(D,Q)$ (which is the uniform source in many cases, in analogy to our
continuous--valued uniform source $U$). For this source,
\begin{equation}
F(D,R)=\left\{\begin{array}{ll}
0 & R\le R(D,Q^*)\\
\infty & R > R(D,Q^*)\end{array}\right.
\end{equation}
This is the case where the entire source space can be fully covered by
spheres of noramlized radius $D$.
In this case, the minimization range in the expression of $e(D)$
obviously reduces to the range $R\le R(D,Q^*)$, 
where the
contribution of the source coding exponent vanishes and hence we are left with 
$e(D)=E[R(D,Q^*)]$. This can be seen as follows:
\begin{eqnarray}
e(D)&=&\min_R[F(D,R)+E(R)]\nonumber\\
&=&\min_{R\le R(D,Q^*)}[0+E(R)]\nonumber\\
&=&E[R(D,Q^*)].
\end{eqnarray}
We now argue that this is a case where separate source-- and channel coding
happens to be optimal: If the source sequence space is fully covered by
spheres of radius $D$,
the source encoder contributes nothing to the excess distortion
event and so, excess distortion may happen only in the event of a channel
error whose exponent is $E(R)$, computed at $R=R(D,Q^*)$, 
which is exactly the above
mentioned expression of $e(D)$. Indeed, from the mathematical point of view, 
the source--channel excess
distortion exponent pertaining to separate source- and channel coding,
denoted by $e_{sep}(D)$, and given by 
$\sup_R\min\{ F(D,R),E(R)\}$, is also equal to
$E[R(D,Q^*)]$ in this case. This is easily shown as follows:
\begin{eqnarray}
e_{sep}(D)&=&\sup_R\min\{F(D,R),E(R)\}\nonumber\\
&=&\sup_R\left\{\begin{array}{ll}
0 & R\le R(D,Q^*)\\
E(R) & R> R(D,Q^*)\end{array}\right.\nonumber\\
&=&E[R(D,Q^*)].
\end{eqnarray}
This is clearly analogous to our case: We fully
cover the unit interval with small intervals of size 
$2e^{-RT}$ using 
a rate--$R$ source code. Similarly, in the $d$--dimensional 
case to be described in Subsection \ref{md}, we perfectly
cover the unit cube by boxes 
of sizes $2e^{-R_1T}\times\ldots\times 2e^{-R_dT}$
using a code of rate $R_1+\ldots+R_d$.

\section{Extensions}

In this section, we extend Theorems 1 and 2 in several directions (one at a
time). These
include the multidimensional case, more general channels, and allowing
a variable power that depends on the parameter.

\subsection{The Multidimensional Case}
\label{md}

The extension to a multidimensional
parameter vector is conceptually quite straightforward.
Suppose now that the parameter is a vector
$\bu=(u_1,\ldots,u_d)\in[-1/2,+1/2)^d$, which is a
realization of a random vector $\bU=(U_1,\ldots,U_d)$, uniformly distributed
over the $d$--dimensional unit hypercube $[-1/2,+1/2)^d$.
Consider now the probability
$$\mbox{Pr}\left[\bigcup_{i=1}^d
\left\{|\hat{U}_i-U_i|> e^{-R_iT}\right\}\right].$$
Then, here
both in the upper bound and the lower bound,
the $d$--dimensional unit cube is divided
by a Cartesian grid with about $e^{R_iT}$ points in each dimension,
$i=1,2,\ldots,d$, thus a total
of $e^{(R_1+R_2+\ldots+R_d)T}$ points,
which means an effective rate of $R_1+R_2+\ldots+R_d$.
More precisely, the lower bound is now
given by
\begin{equation}
\mbox{Pr}\left[\bigcup_{i=1}^d
\left\{|\hat{U}_i-U_i|> e^{-R_iT}\right\}\right]\ge
(1-e^{-\epsilon T})^d\exp\{-T[E(R_1+R_2+\ldots+R_d-\epsilon d)+
o(T)]\} 
\end{equation}
since the
integration in eq.\ (\ref{zz}) now becomes $d$--dimensional.
In the upper bound, we are again quantizing and transmitting one of
$e^{(R_1+R_2+\ldots+R_d)T}$
orthogonal codewords, the one which represents the corresponding
quantization cell. Thus, the probability of the undesired event in question is
of the exponential order
of $e^{-E(R_1+R_2+\ldots+R_d)T}$. Considering the case $R_i=R$ for all
$i\in\{1,2,\ldots,d\}$ (hence $\sum_iR_i=R\cdot d$),
there is then an interesting threshold effect in the
dimensionality of the problem: For $R=0$ (in the weak sense), the exponential
rate of decay of the
probability of the large deviations
event $\cup_{i=1}^d\{|\hat{U}_i-U_i|\ge e^{-0\cdot T}\}$ is
essentially $E(0)=C/2$, independently of $d$. For $R>0$,
the behavior is as follows:
As long as
\begin{equation}
d < d_c\eqdf \lfloor C/R\rfloor,
\end{equation}
the probability of the event
$\cup_{i=1}^d\{|\hat{U}_i-U_i|\ge e^{-RT}\}$
tends to zero as $T\to\infty$. But when $d$ exceeds
$d_c$ and hence the effective rate $R\cdot d$
exceeds $C$, the
probability tends to unity. Thus, $d_c$ is a critical dimension
in this sense. This abrupt transition from $0$ to $1$ in the limiting
probability of excess error is another aspect of the threshold effect.
In most estimation problems we normally encounter, the estimation
performance degrades with the dimensionality (an effect known as the
``curse of dimensionality''),
but usually the degradation is
graceful and not abrupt as here.

All this discussion can be extended, in principle, from Cartesian lattices
in the parameter space to general lattices, where the
undesired excess error event is defined as
the event where the estimated parameter vector
falls outside the respective Voronoi cell centered at the true parameter vector.
Here, the effective rate to be used as the argument of the reliability
function is determined by the normalized logarithm of the
ratio between the volume of the source vector space and the volume of a basic
cell.

\subsection{Other Channels}

The assumption of an AWGN channel with unlimited bandwidth
was not used very strongly beyond the fact that for this particular channel, the
reliability function is fully known for the 
entire range of rates, $0\le R\le
C$. But the reliability function is also known for the Poisson channel
with unlimited bandwidth \cite{Wyner88a}, \cite{Wyner88b}. Here too, the
idea would be to first quantize the parameter and then to use a good code for
the Poisson channel, that asymptotically achieves the reliability function,
e.g., the Wyner code (see also \cite{NVW12}).
Similar comments apply also to more general channels in the limit
of the infinite bandwidth regime \cite{Gallager88}.

In the discrete--time case, the reliability function may not be
known for the entire range of rates, but it is known for all rates above
the critical rate, where it is also achievable by random coding. Moreover,
even if the channel is not fully known, we can derive a universal estimator
that relies on a universal decoder for memoryless channels (see, e.g.,
\cite{FM02} and references therein), on the basis
of the proof of the achievability in Theorem 2. But even at rates below the
critical rate, where the reliability function is not known, the basic
principle of optimum modulation--estimation using a separation--based
scheme continues to hold: First quantize $U$ uniformly and then apply an
optimum channel code.

The modification of our results to discrete memoryless channels also
enables to handle, at least partially, the case of the AWGN channel
with limited bandwidth. This is because
the case of limitation to finite bandwidth $W$ 
is asymptotically equivalent to the discrete
memoryless Gaussian channel with $N=2WT$ channel uses (pertaining to $N=2WT$
orthonormal basis functions that span the subspace of allowable signals).
In this case, $E(R)$ for high rates agrees with the sphere--packing bound,
which in the Gaussian band--limited case is given by
\begin{equation}
E_{sp}(R)=\max_{\rho\ge 0}\left\{\rho
W\ln\left[1+\frac{S}{N_0W(1+\rho)}\right]-\rho R\right\}.
\end{equation}
The critical rate beyond which $E_{sp}(R)=E(R)$ is given by
\begin{eqnarray}
R_c(W)&=&\frac{\partial}{\partial\rho}\left\{\rho
W\ln\left[1+\frac{S}{N_0W(1+\rho)}\right]\right\}\bigg|_{\rho=1}\nonumber\\
&=&W\left[\ln\left(1+\frac{S}{2N_0W}\right)-\frac{1}{2}\cdot\frac{S}{S+2N_0W}\right],
\end{eqnarray}
where the maximum over $\rho$ in achieved within the interval $[0,1]$.

\subsection{The AWGN Channel With Rayleigh Fading}

Another important channel model is the 
AWGN channel with Rayleigh fading.
Here, the signal model is
\begin{equation}
y(t)=a\cdot x(t,u)+n(t),~~~~t\in [0,T)
\end{equation}
where $a$ is a realization 
of a Rayleigh random variable $A$, whose pdf is given by
\begin{equation}\label{e1}
f_A(a)=\frac{a}{\sigma^2}e^{-\frac{a^2}{2 \sigma^2}}, \hspace{3ex} a \geq 0.
\end{equation}
It is assumed that $A$ is independent of $U$, as well as of the noise
$\{n(t),~0\le t< T\}$.
It is instructive to examine the best achievable behavior of the probability of
excess estimation error under this fading model. 

For a given $A=a$, the received signal has power $a^2
S$, which implies that the 
channel capacity is $a^2S/N_0=a^2 C$. 
Correspondingly,
the reliability function
is given by
\begin{equation}
\label{ERa1}
E_a(R)=\left\{\begin{array}{ll}
a^2\frac{C}{2}-R, & 0\le R\le a^2 \frac{ C}{4}\\
(a \sqrt{C}-\sqrt{R})^2, & a^2 \frac{ C}{4}\le R \le a^2 C \\
0, & R \geq a^2 C \end{array} \right.
\end{equation}
Equivalently, if we think of $E_a(R)$ as a function of $a$ parametrized by
$R$, then
\begin{equation}
\label{ERa2}
E_a(R)=\left\{\begin{array}{ll}
a^2 \frac{C}{2}-R, &  a \geq 2 \sqrt{\frac{R}{C}}\\
(a \sqrt{C}-\sqrt{R})^2, & \sqrt{\frac{R}{C}} \le a \le 2 \sqrt{\frac{R}{C}}
\\
0, & a \leq \sqrt{\frac{R}{C}} \end{array} \right.
\end{equation}
In view of Theorems 1 and 2,
averaging the upper and lower bounds on the 
probability of decoding error
given $a$, would yield respective bounds for the fading channel.
For the lower bound, this 
averaging is legitimate
as it corresponds to a receiver that is informed of the realization $a$
of the random variable $A$. For the upper bound, this is legitimate too since
the ML decoder does not depend on (the possibly unknown value of) $a$ 
in the regime of equal--energy signals considered here.

As before, one should distinguish between the cases $R>0$ and $R=0$
(in the strong sense). The following two results are shown in Appendix A.
For the case $R> 0$, the probability of excess estimation error is
essentially equal (for large $T$) to the probability of channel outage,
which is
\begin{equation}
\mbox{Pr}\{A\le \sqrt{R/C}\}=1-e^{-R/2\bar{C}},
\end{equation}
where $\bar{C}=\sigma^2C$ designates the average capacity of the channel.
In other words, there is no decay as $T\to \infty$.
For $R=0$, the best achievable probability of excess estimation error
decays at the rate of $1/T$ rather than exponentially with $T$.

\subsection{Variable Transmission Power}
\label{varpower}

In Section 2, we have restricted the class of modulators in a manner that the
power of the transmitted signal, $\{x(t,u),~0\le t < T\}$, is always $S$,
independently of $u$.
Consider the somewhat broader setting, where the power of $\{x(t,u),~0\le t <
T\}$, denoted $S(u)$, is allowed to depend on $u$, and we only limit the
average power according to
\begin{equation}
\label{avgpower}
\bE\{S(U)\}=\int_{-1/2}^{+1/2}\mbox{d}u\cdot S(u)\le S.
\end{equation}
We argue that our results apply to this wider class of modulators as
well. 

Concerning the achievability, we continue to use the same modulator and
estimator as in the proof of Theorem 2, where the power is $S(u)=S$
for every $u$. The proof of Theorem 1, on the other
hand, has to be
extended to allow variable power. The point is that the proof of Theorem 1
in Section 2 relies heavily on the lower 
bound on the probability of error in $M$--ary
signal detection, which in \cite[Section 3.6.1]{VO79}, is derived under the
assumption of equal--energy signals, 
and we are not aware of an existing extension of
this result to allow sets of signals 
with different energies, where the limitation
is on the average energy only. In Appendix B, we extend the proof of
Theorem 1 to accommodate a given 
average energy constraint, or equivalently, an average
power constraint (\ref{avgpower}). In a nutshell, the intuition 
is that when some of the signals have higher power and some have lower power,
the probability of
error is basically dominated by the those with the lower power, which is,
of course, smaller than the average $S$. Thus, variable power signal sets
offer no improvement relative to fixed power signal sets in terms
of achievable error exponents.

\section*{Acknowledgments}
I would like to thank Yariv Ephraim for many useful discussions and comments
in the course of this work.
Interesting discussions with Tsachy Weissman and Yonina 
Eldar are also acknowledged with thanks.

\section*{Appendix A}
\renewcommand{\theequation}{A.\arabic{equation}}
    \setcounter{equation}{0}

In this appendix, we derive the results for the fading channel for the case
$R> 0$ and the case $R=0$.

Consider the case $R>0$ first. Here,
there is a positive probability that
$a$ would be small enough that the corresponding capacity $a^2C$
would fall below the given $R$, which is exactly the event of channel outage.
This happens with probability
\begin{equation}
\mbox{Pr}\{A^2C < R\}=\int_0^{\sqrt{R/C}}\mbox{d}a\cdot\frac{a}{\sigma^2}
e^{-a^2/2\sigma^2}=1-e^{-R/2\sigma^2 C}=1-e^{-R/2\bar{C}}.
\end{equation}
Owing to the discussion in Subsection 3.1, in the event of outage, the
probability of excess estimation error is very close to unity, and so,
the overall probability of excess estimation error is essentially lower
bounded by the outage probability, i.e.,
\begin{equation}
\mbox{Pr}\{|\hat{U}-U|> e^{-RT}\} \ge [1-o(T)]\cdot (1-e^{-R/2\bar{C}}),
\end{equation}
that is, the probability of excess estimation error no longer decays as $T$
grows without bound.
Concerning the upper bound, we have from eq.\ (\ref{ERa2})
\begin{eqnarray}
\mbox{Pr}\{|\hat{U}-U|> e^{-RT}\}&\le&\int_0^\infty\mbox{d}a
\frac{a}{\sigma^2}e^{-a^2/2\sigma^2}e^{-TE_a(R)}\nonumber\\
&=&\int_0^{\sqrt{R/C}}\mbox{d}a\cdot\frac{a}{\sigma^2}
e^{-a^2/2\sigma^2}+\nonumber\\
& &\int_{\sqrt{R/C}}^{2\sqrt{R/C}}\mbox{d}a\cdot\frac{a}{\sigma^2}
e^{-a^2/2\sigma^2}e^{-T(a\sqrt{C}-\sqrt{R})^2}+\nonumber\\
& &\int_{2\sqrt{R/C}}^{\infty}\mbox{d}a\cdot\frac{a}{\sigma^2}
e^{-a^2/2\sigma^2}e^{-T(a^2C/2-R)}\nonumber\\
&=& 1-e^{-R/2\bar{C}}+o(T),
\end{eqnarray}
where the last line follows from the fact that the
above two last integrals, over the ranges $[\sqrt{R/C},2\sqrt{R/C})$ and
$[2\sqrt{R/C},\infty)$,
both vanish as $T\to\infty$, as can easily be shown.
Thus, the lower bound and the upper bound asymptotically coincide.

As for the case $R=0$ (i.e., $\Delta$ fixed, but small), then in view of
the derivations in Section 2 (see eqs.\ (\ref{lr=0}) and (\ref{ur=0})), for a given $a$,
both the upper bound and the lower bound on the probability
of excess estimation error probability admit
the form $\alpha\cdot Q(a\sqrt{CT\beta})$, where $\alpha$ and $\beta$ are
constants. The respective constants,
$\alpha$ and $\beta$,
pertaining to the upper bound
and the lower bound, are different. However, $\alpha$ is just a multiplicative
constant, which is of secondary importance here, because we are primarily
interested in the rate of decay of both bounds as $T\to \infty$. On the
other hand, $\beta$ is very close to unity in both bounds when
$\Delta$ is small. Thus, the quantity of interest is basically
the expectation of $Q(A\sqrt{CT})$ w.r.t.\ the randomness of $A$.
We next show that for large $T$, 
this quantity is well--approximated by
\begin{equation}
\bE\{Q(A\sqrt{CT})\}=\int_0^\infty
\mbox{d}a
\frac{a}{\sigma^2}e^{-a^2/2\sigma^2}Q(a\sqrt{CT})\approx \frac{1}{4\bar{C}T},
\end{equation}
that is, the minimum achievable excess estimation error probability decays
algebraically rather than exponentially.
Using Craig's formula (see, e.g., \cite{TA99}),
\begin{equation}
Q(x)=\frac{1}{\pi}\int_0^{\pi/2}\mbox{d}\theta
\exp\left(-\frac{x^2}{2\sin^2\theta}\right),
\end{equation}
we have the following:
\begin{eqnarray}
\bE\{Q(A\cdot\sqrt{CT})\}&=&
\int_0^\infty\frac{\mbox{d}a}{\sigma^2}\cdot ae^{-a^2/2\sigma^2}
Q(a\cdot\sqrt{CT})\nonumber\\
&=&
\int_0^\infty\frac{\mbox{d}a}{\sigma^2}\cdot ae^{-a^2/2\sigma^2}\cdot
\frac{1}{\pi}\int_0^{\pi/2}\mbox{d}\theta\exp\left(-\frac{a^2CT}{2\sin^2\theta}\right)\nonumber\\
&=&
\frac{1}{\pi}\int_0^{\pi/2}\mbox{d}\theta
\int_0^\infty\mbox{d}\left(\frac{a^2}{2\sigma^2}\right)
\exp\left(-\frac{a^2}{2\sigma^2}\left[1+\frac{\bar{C}T}{\sin^2\theta}\right]\right)\nonumber\\
&=&
\frac{1}{\pi}\int_0^{\pi/2}\frac{\mbox{d}\theta}{1+\bar{C}T/\sin^2\theta}\nonumber\\
&=&
\frac{1}{\pi}\int_0^{\pi/2}\frac{\mbox{d}\theta\cdot\sin^2\theta}{\bar{C}T+\sin^2\theta}.
\end{eqnarray}
The last expression can be upper bounded and lower bounded by
bounding the $\sin^2\theta$ term of the denominator by $0$ and $1$,
respectively. Both bounds are well approximated by 
$1/(4\bar{C}T)$ for large $T$.

\section*{Appendix B}
\renewcommand{\theequation}{B.\arabic{equation}}
    \setcounter{equation}{0}

In this appendix, we provide an outline of the 
extension of Theorem 1 to the variable power case.

For a given $u$ and $\Delta$, consider again the grid
$\{u+i\Delta\}_{i=0}^{M-1}$, which is assumed to lie entirely in
$[-1/2,+1/2)$, and let $\Delta=2e^{-RT}$ and $M=e^{(R-\epsilon)T}/2+1$, as
before. Let $S_{\min}\eqdf\min_uS(u)$ and $S_{\max}\eqdf\max_uS(u)$. We first argue
that for modulators whose power function $S(u)$ is 
continuous (or at least, left- or right--continuous) in the vicinity
of its minimum, the assertion of
Theorem 1 is rather straightforward in the range $C_{\min} \le R
< C$, where $C_{\min}=S_{\min}/N_0$. The reason is that the grid points,
$\{u+i\Delta\}_{i=0}^{M-1}$, where $u$
is near the minimum of the power
function (and so are all other grid points, with the above
assignment of $M$ and $\Delta$), constitute a signal set whose
rate, $R-\epsilon$, is very close to (or even exceeds)
its capacity, which is about $C_{\min}$, since
all signals in this grid have power near $S_{\min}$. Thus, this grid dominates
the probability of error (and hence also the probability of 
excess estimation error) and
it dictates a sub-exponential decay at best, which
is trivially lower bounded by the exponent $\exp[-TE(R)]$. 
In view of this, we shall confine attention throughout to the range of rates
$0< R < C_{\min}$. 

Now, consider
the partition of the range of powers $[S_{\min},S_{\max}]$ into small bins
of width $\delta$, where $\delta$ is assumed to divide $S_{\max}-S_{\min}$. 
For a given $u$, let 
$\calC_i$ denote the subset of integers $\{j\}$ for which
$S(u+j\Delta)$ falls in the $i$--th bin, that is,
\begin{equation}
S_{\min}+i\delta \le S(u+j\Delta) < S_{\min}+(i+1)\delta,~~~~i=0,1,\ldots,r-1
\end{equation}
where $r=(S_{\max}-S_{\min})/\delta$. 
First, observe that
\begin{equation}
P_e(u,\Delta)\ge \sum_{i=0}^{r-1}\frac{|\calC_i|}{M}P_e(\calC_i),
\end{equation}
where $P_e(\calC_i)$ is the probability of error pertaining to the subset of
signals $\{x(t,u+j\Delta)\}_{j\in \calC_i}$ alone. The reason for the inequality is that
error events associated with
confusion between pairs of signals that belong to different bins are not counted
in the r.h.s.
Next, for a given $\epsilon > 0$,
let $\calI_\epsilon$ denote the index set 
$\{i:~|\calC_i|\ge e^{-\epsilon T}M\}$.
Then, obviously, $P_e(u,\Delta)$ is further lower bounded by
\begin{equation}
P_e(u,\Delta)\ge \sum_{i\in \calI_\epsilon}\frac{|\calC_i|}{M}P_e(\calC_i).
\end{equation}
Now, let us slightly alter the powers of all signals in $\calC_i$ to be
$S_i\eqdf S_{\min}+(i+1/2)\delta$, 
neglecting the effect that this may have on the
exponent of $P_e(\calC_i)$.\footnote{The lower bounds on the probability of
error of $M$ equal--energy signals are straightforwardly extended to allow
almost equal powers (within $\pm\delta/2$), with only a small degradation in the
exponential rate,
which depends on $\delta$.}
Let us denote here the reliability function of a
rate--$R$ code with power $S$ by $E(R,S)$, to emphasize the dependence on the
power (via the dependence on the capacity). Then, for every
$i\in\calI_\epsilon$, we have
\begin{equation}
P_e(\calC_i)\ge e^{-T[E(R-2\epsilon,S_i)+o(T)]},
\end{equation}
since the size of $\calC_i$ is of the exponential order of at least
$e^{(R-2\epsilon)T}$.
Also, let us denote
\begin{equation}
\pi_i=\frac{|\calC_i|}{\sum_{j\in\calI_\epsilon}|\calC_j|}.
\end{equation}
Then,
\begin{equation}
\label{factors}
P_e(u,\Delta)\ge \sum_{i\in
\calI_\epsilon}\frac{|\calC_i|}{M}\cdot
\sum_{i\in\calI_\epsilon}\pi_ie^{-T[E(R-2\epsilon,S_i)+o(T)]}.
\end{equation}
As for the first factor on the r.h.s.\ of (\ref{factors}), we have 
\begin{equation}
1=\sum_{i\in \calI_\epsilon}\frac{|\calC_i|}{M}+
\sum_{i\in \calI_\epsilon^c}\frac{|\calC_i|}{M}\le
\sum_{i\in \calI_\epsilon}\frac{|\calC_i|}{M}+re^{-\epsilon T},
\end{equation}
and so, this factor is lower bounded by $(1-re^{-\epsilon T})$.
Now, observe that the function $e^{-TE(R-2\epsilon,S)}$
is convex\footnote{The function $e^{-Tf(x)}$ is convex in $x\in \calX$ 
whenever $f$ is twice differentiable and
$T\ge \sup_{x\in\calX}f''(x)/|f'(x)|^2$,
as can easily be seen from the
second derivative of $e^{-Tf(x)}$. An alternative consideration is 
that for large $T$, the average
of $e^{-Tf(x)}$ is dominated by $e^{-T\inf_{x\in\calX}f(x)}$, 
and that $\inf_{x\in\calX}f(x)$ is
smaller than the average of $f(x)$ over $\calX$.}
in $S$ for all $T> \sqrt{R}/[2\sqrt{C_{\min}}(\sqrt{C_{\min}}-\sqrt{R})^2]$. 
It follows then from
(\ref{factors}) that
\begin{equation}
\label{1stbound}
P_e(u,\Delta)\ge (1-re^{-\epsilon T})\cdot
\exp\left[-TE\left(R-2\epsilon,\sum_{i\in\calI_\epsilon}\pi_iS_i\right)+o(T)\right].
\end{equation}
Next, we need an upper bound on $\sum_{i\in\calI_\epsilon}\pi_iS_i$. 
This is accomplished as follows:
\begin{eqnarray}
\bar{S}(u)&\eqdf&\frac{1}{M}\sum_{i=0}^{M-1}S(u+i\Delta)\nonumber\\
&\ge&\sum_{i\in\calI_\epsilon}\frac{|\calC_i|}{M}\cdot S_i+
\sum_{i\in\calI_\epsilon^c}\frac{|\calC_i|}{M}\cdot S_i-\frac{\delta}{2}\nonumber\\
&\ge&\sum_{i\in\calI_\epsilon}\frac{|\calC_i|}{M}\cdot S_i-\frac{\delta}{2}
\end{eqnarray}
and so,
\begin{equation}
\sum_{i\in\calI_\epsilon}\pi_iS_i\le
\frac{\bar{S}(u)+\delta/2}{\sum_{i\in\calI_\epsilon}|\calC_i|/M}\le
\frac{\bar{S}(u)+\delta/2}{1-re^{-\epsilon T}}.
\end{equation}
Thus, from (\ref{1stbound}), we have
\begin{equation}
P_e(u,\Delta)\ge (1-re^{-\epsilon T})\cdot
\exp\left[-TE\left(R-2\epsilon,\frac{\bar{S}(u)+\delta/2}{1-re^{-\epsilon T}}\right)+o(T)\right].
\end{equation}
Finally, we integrate both sides of the last inequality w.r.t.\ $u$, in order
to relate it to the probability of excess estimation error, as in the proof of
Theorem 1. To this end, we first observe the following:
\begin{eqnarray}
\int_{-1/2}^{1/2-(M-1)\Delta}\mbox{d}u\cdot\bar{S}(u)&=&
\int_{-1/2}^{1/2-(M-1)\Delta}\mbox{d}u\cdot\frac{1}{M}\sum_{i=0}^{M-1}S(u+i\Delta)\nonumber\\
&=&\frac{1}{M}\sum_{i=0}^{M-1}\int_{-1/2}^{1/2-(M-1)\Delta}\mbox{d}u\cdot S(u+i\Delta)\nonumber\\
&=&\frac{1}{M}\sum_{i=0}^{M-1}
\int_{-1/2+i\Delta}^{1/2-(M-1)\Delta+i\Delta}\mbox{d}u\cdot S(u)\nonumber\\
&\le&\frac{1}{M}\sum_{i=0}^{M-1}
\int_{-1/2}^{1/2}\mbox{d}u\cdot S(u)\nonumber\\
&=&\int_{-1/2}^{1/2}\mbox{d}u\cdot S(u)\nonumber\\
&\le&S,
\end{eqnarray}
and therefore, for the above defined assignments of $\Delta$ and $M$, we have:
\begin{equation}
\frac{1}{1-e^{-\epsilon T}}\int_{-1/2}^{1/2-e^{-\epsilon T}}\mbox{d}u\cdot\bar{S}(u)\le
\frac{S}{1-e^{-\epsilon T}}.
\end{equation}
Thus,
\begin{eqnarray}
& &\mbox{Pr}\{|\hat{U}-U|> e^{-RT}\}\nonumber\\
&\ge&\int_{-1/2}^{1/2-e^{-\epsilon T}}\mbox{d}u\cdot P_e(u,2e^{-RT})\nonumber\\
&\ge&(1-re^{-\epsilon T})(1-e^{-\epsilon
T})\int_{-1/2}^{1/2-e^{-\epsilon T}}\frac{\mbox{d}u}{1-e^{-\epsilon T}}\cdot
\exp\left[-TE\left(R-2\epsilon,\frac{\bar{S}(u)+\delta/2}{1-re^{-\epsilon T}}\right)+o(T)\right]
\nonumber\\
&\ge&(1-re^{-\epsilon T})^2
\exp\left[-TE\left(R-2\epsilon,\frac{S+\delta/2}{(1-re^{-\epsilon
T})(1-e^{-\epsilon T})}\right)+o(T)\right]\nonumber\\
&=&(1-re^{-\epsilon T})^2e^{-T[E(R,S)+o'(T)]},
\end{eqnarray}
where in the last line, $o'(T)$ means another function (other than
$o(T)$ of the previous lines) that tends to $0$ as $T\to\infty$,
which is obtained by letting $\epsilon$ and $\delta$ tend to zero
as $T\to\infty$ at the appropriate rates (e.g.,
$\epsilon=\delta=1/\sqrt{T}$).

%\section*{Appendix}
%\renewcommand{\theequation}{A.\arabic{equation}}
%    \setcounter{equation}{0}


\begin{thebibliography}{AA}

\bibitem{BSET97}
K.~L.~Bell, Y.~Steinberg, Y.~Ephraim, and H.~L.~van Trees, ``Extended
Ziv---Zakai lower bounds for vector parameter estimation,'' {\it IEEE Trans.\ Inform.\
Theory}, vol.\ 43, no.\ 2, pp.\ 626--637, March 1997.

\bibitem{BE09}
Z.~Ben--Haim and Y.~C.~Eldar, ``A lower bound on the Bayesian MSE
based on the optimal bias function,''  
{\em IEEE Trans.~Inform.~Theory\/},
vol.\ 55, no.~11, pp.~5179--5196, November 2009.

\bibitem{CZZ75}
D.~Chazan, M.~Zakai, and J.~Ziv, ``Improved lower bounds on %PE-069
signal parameter estimation,''
{\it IEEE Trans.\ Inform.\ Theory}, vol.\ IT--21, no.\ 1, pp.\ 90--93,
January 1975.

\bibitem{Csiszar80}
I.~Csisz\'ar, ``Joint source--channel error exponent,'' {\it Problems of
Control and Information Theory}, vol.\ 9, no.\ 5, pp.\ 315--328, 1980.

\bibitem{Csiszar82}
I. Csisz\'ar, ``On the error exponent of source-channel transmission with
a distortion threshold,''
{\em IEEE Trans.~Inform.~Theory\/},
vol.~IT--28, no.~6, pp.~823--828, November 1982.

\bibitem{FM02}
M.~Feder and N.~Merhav, ``Universal composite hypothesis
testing: a competitive minimax approach,''
{\it IEEE Trans.\ Inform.\ Theory},
special issue in memory of Aaron D.~Wyner,
vol.\ 48, no.\ 6, pp.\ 1504--1517, June 2002.

\bibitem{Gallager68}
R.~G.~Gallager, {\it Information Theory 
and Reliable Communication}, New York, Wiley 1968.

\bibitem{Gallager88}
R.~G.~Gallager, ``Energy limited channels: coding, multiaccess, and %CCTT-082
spread spectrum,'' LIDS Report, LIDS--P--1714, M.I.T., November 1988.

\bibitem{Gray90}
R.~M.~Gray, {\it Source Coding Theory}, Kluwer Academic Publishers, 1990.

\bibitem{KK86}
A.~D.~M.~Kester and W.~C.~M.~Kallenberg, ``Large deviations of %LD-010
estimators,'' {\em Ann.\ Statist.\/}, vol.~14, no.~2, pp.~848--664, 1986.

\bibitem{Lehmann83}
E.~L.~Lehmann, {\it Theory of Point Estimation}, John Wiley \& Sons,
1983.

\bibitem{Marton74}
K.~Marton, ``Error exponent for source coding with a fidelity %RDT-004
criterion, '' {\em IEEE Trans.~Inform.~Theory\/},
vol.~IT--20, no.~2, pp.~197--199, March 1974.

\bibitem{p137}
N.~Merhav, ``Threshold effects in parameter estimation
as phase transitions in statistical mechanics,''
{\it IEEE Trans.\ Inform.\ Theory},
vol.\ 57, no.\ 10, pp.\ 7000--7010, October 2011.

\bibitem{p146}
N.~Merhav, ``Data processing inequalities based on a certain structured class
of information measures with application to estimation theory,''
submitted to {\it IEEE Trans.\ Inform.\ Theory}, September 2011.
[http://arxiv.org/pdf/1109.5351.pdf]

\bibitem{NVW12}
A.~No, K.~Venkat, and T.~Weissman, ``Joint source--channel coding of one
random variable over the Poisson channel,'' submitted to {\it ISIT 2012},
February 2012.

\bibitem{Sherman58}
S.~Sherman, ``Non-mean-square error criteria,'' %FSE-032
{\em IRE Trans. Inform. Theory\/}, pp.~125--126, September 1958.

\bibitem{TA99}
C.~Tellambura and A.~Annamalai, ``Derivation of Craig's formula for Gaussian
probability function,'' {\it Electronic Letters},
vol.\ 35, no.\ 17, pp.\ 1424--1425, August 19, 1999.

\bibitem{vanTrees68}
H.~van Trees, {\it Detection, Estimation, and Modulation Theory}, Part I,
New York, John Wiley \& Sons, 1968.

\bibitem{VO79}
A.~J.~Viterbi and J.~K.~Omura, {\it Principles of Digital Communication and
Coding}, McGraw--Hill, 1979.

\bibitem{Weiss85}
A.~J.~Weiss, {\it Fundamental Bounds in Parameter Estimation},
Ph.D.\ dissertation, Tel Aviv University, Tel Aviv, Israel, June 1985.

\bibitem{WJ65}
J.~M.~Wozencraft and I.~M.~Jacobs, {\it Principles of Communication
Engineering}, John Wiley \& Sons, 1965. Reissued by Waveland Press, 1990.

\bibitem{Wyner88a}
A.~D.~Wyner, ``Capacity and error exponent for the direct detection
photon channel -- part I,''
{\it IEEE Trans.\ Inform.\ Theory}, vol.\ 34, no.\ 6, pp.\ 1449--1461,
November 1988.

\bibitem{Wyner88b}
A.~D.~Wyner, ``Capacity and error exponent for the direct detection
photon channel -- part II,''
{\it IEEE Trans.\ Inform.\ Theory}, vol.\ 34, no.\ 6, pp.\ 1462--1471,
November 1988.

\bibitem{WZ69}
A.~D.~Wyner and J.~Ziv, ``On communication of analog data from a
bounded source space,''
{\em Bell System Technical Journal\/},
vol.~48, no.~10, pp.~3139--3172, December 1969.


\bibitem{ZZ69a}
M.~Zakai and J.~Ziv, ``On the threshold effect in radar range %PE-066
estimation,'' {\it IEEE Trans.\ Inform.\ Theory}, vol.\ IT--15, pp.\ 167--170,
January 1969.

\bibitem{ZZ75}
M.~Zakai and J.~Ziv, ``A generalization of the rate-distortion theory and
applications,'' in: {\em Information Theory New Trends and Open Problems\/},
edited by G. Longo, Springer-Verlag, 1975, pp.~87--123.

\bibitem{ZZ69b}
J.~Ziv and M.~Zakai, ``Some lower bounds on signal parameter %PE-022
estimation,'' {\em IEEE Transactions on Information Theory\/},
vol.~IT--15, no.~3, pp.~386--391, May 1969.

\bibitem{ZZ73}
J.~Ziv and M.~Zakai, ``On functionals satisfying a data-processing %ST-006
theorem,'' {\em IEEE Trans.~Inform.~Theory\/},
vol.~IT--19, no.~3, pp.~275--283, May 1973.

\end{thebibliography}
\end{document}